%% file: main.tex
\documentclass[conference]{IEEEtran}

\usepackage{amsmath,amssymb,amsfonts}
\usepackage[noend]{algorithmic}
\usepackage{graphicx}
\usepackage{textcomp}
\usepackage{xcolor}
\usepackage{hyperref}
\usepackage{svg}
\usepackage{cite}
\usepackage{amsmath,amssymb,amsfonts}
\usepackage{algorithm,multirow}
\usepackage{textcomp}
\usepackage{xcolor}
\usepackage{graphics}
\usepackage{complexity}
\usepackage{comment}
\usepackage{multicol}
\usepackage[autostyle, english = american]{csquotes}
\MakeOuterQuote{"}
\usepackage{hyperref}
\usepackage{caption}
\usepackage{subcaption}
\usepackage{wrapfig}

\def\BibTeX{{\rm B\kern-.05em{\sc i\kern-.025em b}\kern-.08em
    T\kern-.1667em\lower.7ex\hbox{E}\kern-.125emX}}

\IEEEoverridecommandlockouts
\IEEEpubid{\makebox[\columnwidth]{979-8-3503-2445-7/23/\$31.00~\copyright2023 IEEE \hfill} \hspace{\columnsep}\makebox[\columnwidth]{}}

\newcommand{\sol}{DQuLearn}

\begin{document}

\title{Distributed Quantum Learning with co-Management in a Multi-tenant Quantum System}

\author{
    Anthony D'Onofrio Jr.\textsuperscript{\rm 1},
    Amir Hossain\textsuperscript{\rm 1},
    Lesther Santana\textsuperscript{\rm 1},
    Naseem Machlovi\textsuperscript{\rm 1}, \\
    Samuel Stein\textsuperscript{\rm 2},
    Jinwei Liu\textsuperscript{\rm 3}, 
    Ang Li\textsuperscript{\rm 2},
    and Ying Mao\textsuperscript{\rm 1}\\

    \textsuperscript{\rm 1} Computer and Information Science Department, Fordham University, \\ \{adonofrio10,ahossain20,lsantanacarmona,mmachlovi,ymao41\}@fordham.edu\\
    \textsuperscript{\rm 2}Pacific Northwest National Laboratory (PNNL), \{samuel.stein, ang.li\}@pnnl.gov\\
    \textsuperscript{\rm 3} Florida A\&M University, \ jinwei.liu@famu.edu \\
 }

\maketitle

\IEEEpubidadjcol

\begin{abstract}
The rapid advancement of quantum computing has pushed classical designs into the quantum domain, breaking physical boundaries for computing-intensive and data-hungry applications
Given its immense potential, quantum-based computing systems have attracted increasing attention 
with the hope that some systems may provide a quantum speedup. For example, variational quantum algorithms have been proposed for quantum neural networks to train deep learning models on qubits, achieving promising results.
Existing quantum learning architectures and systems rely on single, monolithic quantum machines with abundant and stable resources, such as qubits. However, fabricating a large, monolithic quantum device is considerably more challenging than producing an array of smaller devices. In this paper, we investigate a distributed quantum system that combines multiple quantum machines into a unified system. We propose \sol, which divides a quantum learning task into multiple subtasks. Each subtask can be executed distributively on individual quantum machines, with the results looping back to classical machines for subsequent training iterations. Additionally, our system supports multiple concurrent clients and dynamically manages their circuits according to the runtime status of quantum workers. Through extensive experiments, we demonstrate that \sol~achieves similar accuracies with significant runtime reduction, by up to 68.7\% and an increase per-second circuit processing speed, by up to 3.99 times,  in a 4-worker multi-tenant setting.

\end{abstract}


\input{introduction}

\input{relatedworks}

\input{solution}
\input{result}

\input{discussion}

\bibliographystyle{abbrv}
\bibliography{main}

\end{document}

%% file: introduction.tex
\section{Introduction}

Over the past decade, deep learning systems and applications have seen remarkable advancements. Innovative algorithms, enhanced computational power, and modern designs have facilitated a wide range of use cases~\cite{haghighat2021sciann, zheng2021black}
Large language models and other deep learning-based applications require considerable computational resources and are typically trained, deployed, and maintained on cloud-based distributed systems. 
Numerous studies have explored methods to optimize systems for deep learning applications~\cite{zhao2020distributed, mao2022differentiate, mao2022elastic}. 
However, in the post-Moore's Law era, the approaching physical limits of semiconductor fabrication and the increasing size of datasets raise concerns about the future of deep learning and its potential constraints. 

At the same time, the rapid advancement of quantum computing and its potentially revolutionary promise has motivated researching quantum machine learning designs. 
The immense potential of quantum-based deep learning architectures and applications has attracted growing interest from industry and academia, hoping that some systems may provide quantum speedup. QuClassi~\cite{stein2022quclassi} and QuGan~\cite{stein2021qugan}, for instance, utilize a variational quantum algorithm structure, inducing parameterized circuits on quantum processors and optimizing on a classical machine. 

Despite rapid progress in the Noisy Intermediate-Scale Quantum (NISQ) era, current quantum systems have not yet achieved practical applicability in real-world scenarios. Existing quantum deep learning models and quantum systems face several challenges. First, most current quantum deep learning proposals rely on single, monolithic quantum machines. However, fabricating a large, monolithic quantum device is significantly more challenging than producing arrays of smaller ones. 
Present quantum computer technologies such as superconducting, trapped ion, and photonic quantum computers demonstrate scaling promise \cite{lekitsch2017blueprint,takeda2019toward,smith2022scaling}. 
However, each technology suffers from both unique and generalized issues such as superconducting qubits being susceptible to decoherence, trapped ion processors being limited in qubit scalability, and gate infidelities challenging all technologies. 
Current quantum machine learning approaches primarily focuses on implementing algorithms with fewer qubits. Consequently highlighting the challenge in leveraging multiple small quantum computers for quantum deep learning models instead of a single, monolithic device, which is left primarily unoccupied. 
Additionally, public quantum computing platforms, such as IBM-Q, use a single-tenant mechanism, where one user occupies the entire machine while others wait in a queue. This design compromises quantum resource efficiency if specific applications do not fully utilize the available qubits. 


To address these challenges, we investigate a distributed quantum learning architecture, \sol, which efficiently utilizes quantum resources from multiple workers across the entire system. \sol~breaks down a large quantum task into smaller subtasks to accommodate low-qubit machines. These subtasks are then distributed to individual workers for execution. The results loop back to classical machines and are analyzed for the next training iteration. This distributed design overcomes the limited resources of a single machine and enables applications to leverage resources across multiple quantum workers. Furthermore, our quantum-classical co-Management modules dynamically manages the circuits from clients and distributes the workload to the most suitable quantum worker based on their runtime status. \sol~has been evaluated with both controlled and uncontrolled computing environments. We summarize the key contributions below.

\begin{itemize}
    \item We propose \sol, a distributed quantum learning architecture that divides a training task into multiple subtasks. These subtasks can be executed on different quantum machines distributively, independent of each other, leveraging resources across the entire system.

    \item We develop a distributed quantum system that combines multiple quantum workers and supports multiple concurrent clients. It accepts computing tasks from clients and selects the most suitable quantum worker based on their runtime status.

    \item We evaluate our system using IBM-Q simulation backends, an uncontrolled cloud environment provided by IBM. Additionally, we conduct experiments on commercial clouds within a controlled, multi-tenant computing environment. The results demonstrate the effectiveness achieved by \sol~along with significant improvements in runtime reduction, by up to 68.7\%, and processing speed, by up to 3.9 times.

\end{itemize}

%% file: relatedworks.tex
\section{Related Works}

Given the ever increasing data and model size in machine learning, distributed learning has been studied aiming to leverage computing resources across an entire cluster~\cite{ben2019demystifying, mayer2020scalable}. 
Researchers developed programming frameworks
and system architectures~\cite{barrachina2021pydtnn, chen2019distributed, wang2019distributed} for distributed and parallel deep learning applications. 
Such optimized designs considerably enhance deep learning performance from a system perspective. 
However, due to ever-growing data sizes, increased complexity of model designs, and hardware limitations, achieving these improvements poses an increasingly challenging task.

With great potential of processing complex problems beyond current capabilities at a fraction of the time, quantum-based algorithms and applications have received great attention recently as innovations in quantum machine learning~\cite{gupta2022comparative, garg2020advances, wilson2021quantum, ruan2022vacsen} have been proposed to improve existing models from various perspectives. 
However, these proposals either focus on theoretical proofs that assume unlimited quantum resources or primitive applications with a toy-like data set. 
A significant hindrance lies in the fact that practical applications of quantum computers require a large number of logical qubits, demanding millions of physical qubits \cite{Martinis2015,Beverland2022}, which is simply unfeasible in the NISQ era. With the current development in quantum frameworks, programming languages and compilers~\cite{aleksandrowicz2019qiskit, broughton2020tensorflow, bichsel2020silq, chong2017programming, liu2021relaxed}, significant speed-ups in prototyping quantum algorithms and applications are attained. Unfortunately, none of the existing systems support distributed quantum computing in a practical setting. 


Due to limited quantum resources, many quantum algorithms employ a quantum-classical design~\cite{brunken2021automated, posenitskiy2021application, stein2021hybrid, mu2022iterative, stein2022qucnn}. For instance, variational quantum algorithms, widely used in quantum chemistry and quantum deep learning, utilize a quantum processor to simulate quantum dynamics and a classical optimizer to improve the results. However, from a system perspective, these applications mainly emphasize the interaction between a single quantum machine and a single classical node, typically lacking parallelization. Furthermore, this single quantum-classical pair system fails to fully exploit a multi-node quantum-classical system. Limited prior works have investigated distributed quantum systems~\cite{feng2022verification, guerreschi2022fast}, but they did not take a quantum-classical system into consideration. 
For example, authors in~\cite{feng2022verification} propose a distributed programming language that allows local quantum operations and classical communication. It investigates the formal description and verification of distributed quantum systems.
Based on established distributed systems, QMPI~\cite{haner2021distributed} proposes an extension of the classical Message Passing Interface to enable high-performance implementations of distributed quantum algorithms. In addition, it includes a quantum communication model (SENDQ) to evaluate the performance of these algorithms and foster algorithmic optimizations. Besides, a distributed quantum simulator, IQS, is presented by Intel Labs~\cite{guerreschi2022fast}. IQS is able to utilize distributed resources in a cloud computing infrastructure and accelerate the simulation. However, they consider a system that consists entirely of quantum computers. The classical computing resources are merely intended for simulation. They fail to leverage these widely-available cloud resources to solve a practical problem with quantum computers collaboratively.

%% file: solution.tex

\section{\sol~ System Design}

In this section, we discuss our system architecture that facilitates distributed quantum learning, the workflow algorithms and management modules of the system in detail.

\subsection{System Architecture}

\begin{figure}[ht]			
	\centering
	\includegraphics[width=0.75\linewidth]{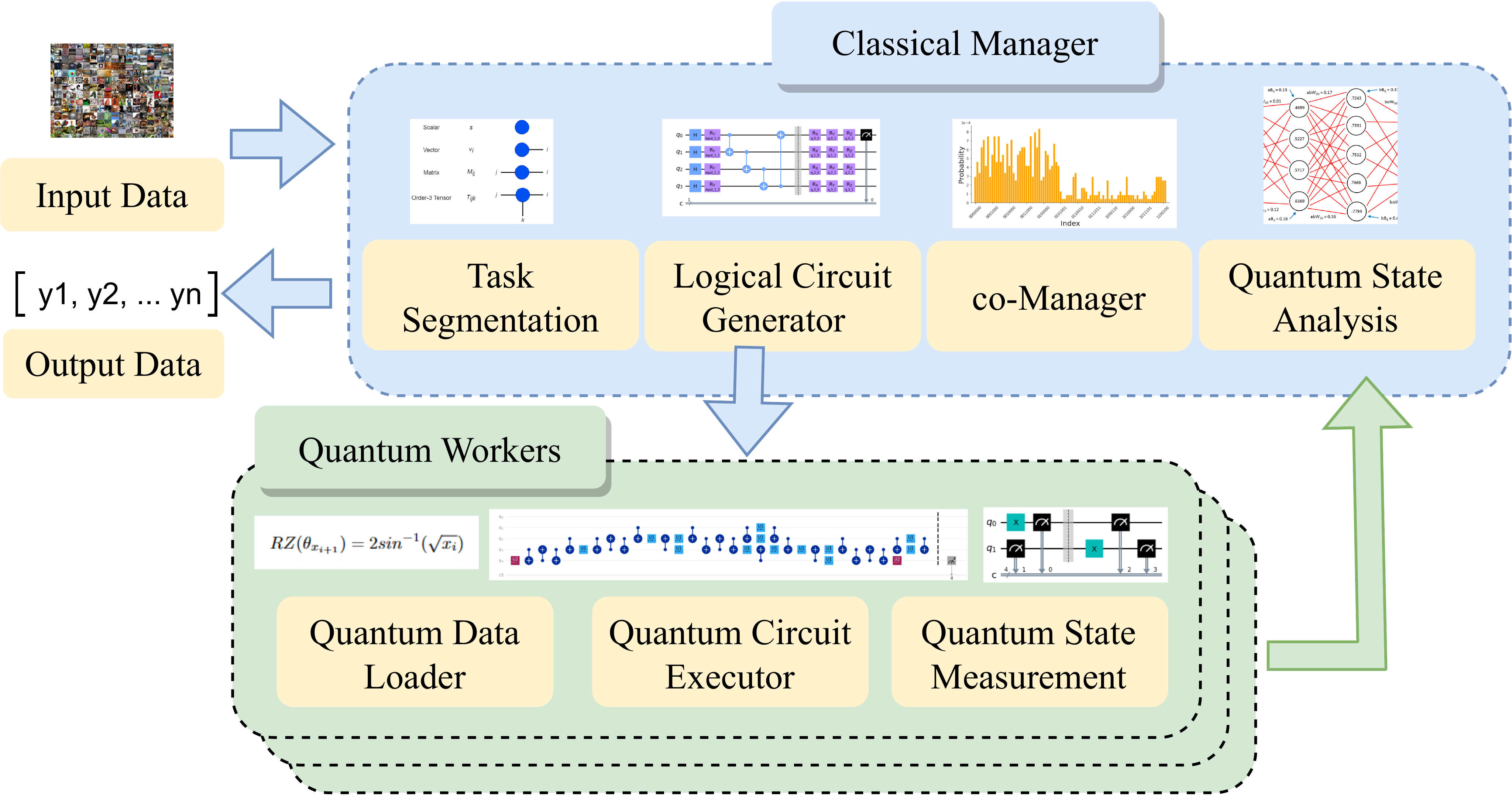}
    \caption{\sol~System Architecture}
    \label{fig:sysem}
\end{figure}
The proposed \sol~ system architecture functions as a closed feedback loop between a classical manager and a set of quantum workers, as illustrated in Figure~\ref{fig:sysem}. The data undergoes an initial cleaning process that includes the removal of significant outliers and other necessary data cleaning procedures. 
The cleaned classical data is then passed to related modules in the system.

\begin{wrapfigure}[10]{l}{0.55\linewidth}
    \centering
    \includegraphics[width=1\linewidth]{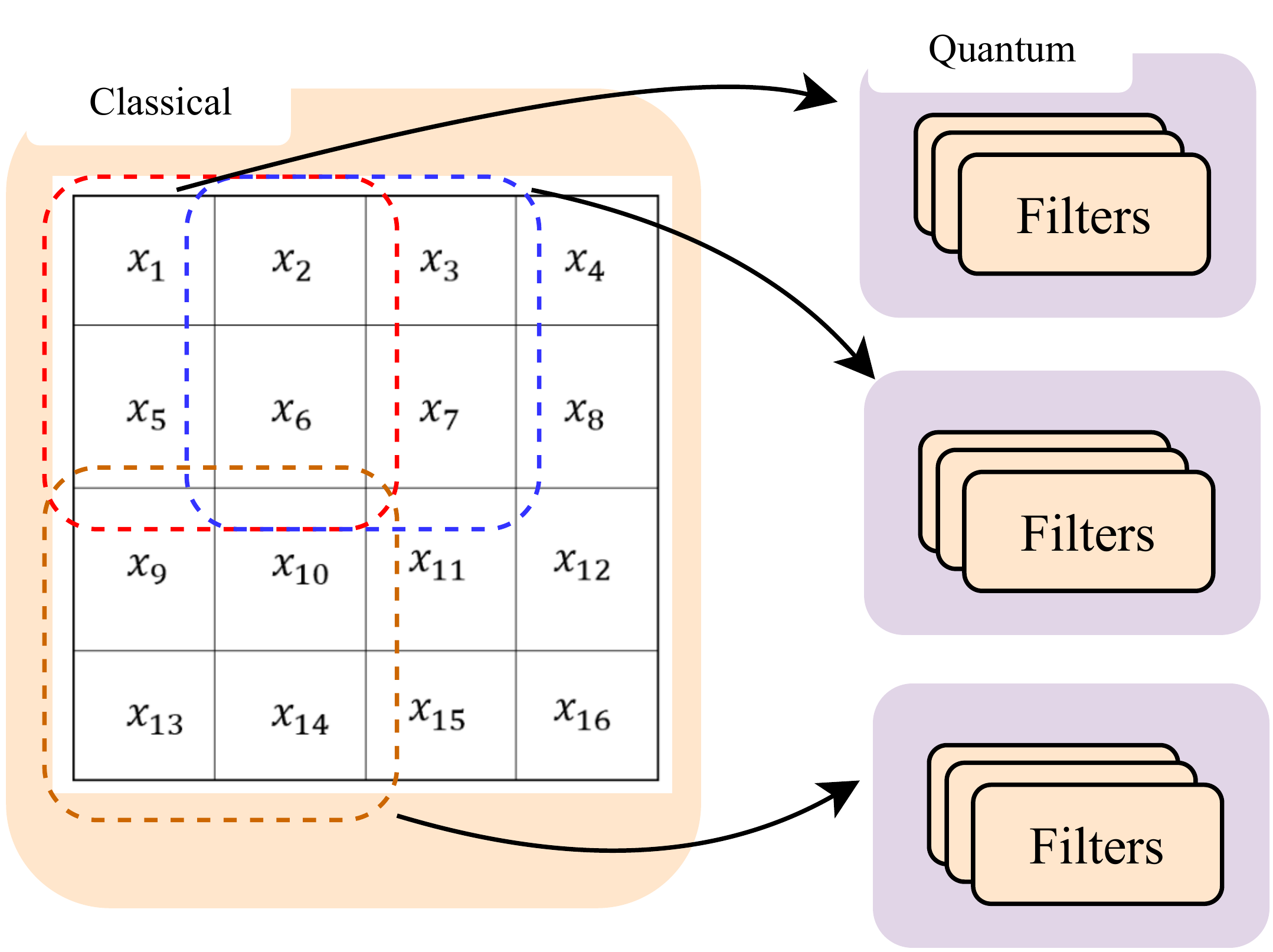}
    \caption{Parallel Filters}
    \label{fig:cnn-filter}
\end{wrapfigure}
The {\em Task Segmentation} is a key module in the system that aims to segment the large input into smaller pieces. Based on the predefined subtask unit, such as convolutional filter size, the Task Segmentation module decomposes the original data into multiple smaller sections in a continued fashion. The data points in the same section can be grouped and treated as individual subtasks.  
Figure~\ref{fig:cnn-filter} illustrates a conceptual example of a convolutional neural network (CNN) for image classification. An image on a classical computer is divided into smaller sections and there might be padding between the sections.

The generated smaller data sections are transferred to the {\em Logical Circuit Generator} module. The main task in this module is to encode the classical data points onto the quantum qubits. Different encoding methods can be adopted and we utilize X and Y rotations to encode our data. After the data transformation, a logical circuit is generated for each section and passes the circuit to next module. Due to limited quantum resources, the original image may be too large to be encoded on quantum Workers, while sub-images are small enough.

The quantum-classical co-Manager module accepts logical circuits from the generator. It executes the scheduling algorithms to monitor a set of quantum computers and selects the best host for each circuit. Once a quantum worker is selected, it distributes the circuit to the targeted worker for execution. 
On a specific quantum worker, it employs three modules to interact with a classical computer. Firstly, the the {\em Quantum Data Loader} is used to load the logical circuits with initial data encoding. It maps logical qubits that are defined by the given circuits onto physial qubits on this specific quantum machine. Next, the {\em Quantum Circuit Executor} conducts the experiment to execute the gates on the circuits. Finally, the {\em Quantum Measurement} calculates the fidelity from one anicilla qubit which is used to calculate model loss, and sends this metric back to the classical computer. 

When the results loop back at classical computer, the {\em Quantum State Analyst} module actively analyzes the data based on the predefined cost function. The module uses this analysis to update the trainable parameters in the quantum circuit in attempts to minimize the cost function. 
In our system, the decomposed data goes through smaller pieces on quantum Workers, and the results are transferred back to the classical computer for further processing. This hybrid quantum-classical design allows computation tasks to be distributed collaboratively. With data stored on multiple classical computers and processed on multiple quantum workers iteratively. Consequently, our system enables a quantum application to be distributed across multiple quantum and classical computers, working concurrently. 

\begin{algorithm}[H]
\caption{\sol~  Algorithm}
\label{alg:cap}
\begin{algorithmic}[1]
\STATE Data set Loading Dataset: $(X \vert Class: Mixed)$
\STATE Parameter Initialization:

\item [] Learning Rate : $\alpha = 10^{-4}$
\item [] Network Weights : $\theta_{d} = [$Rand Num between $0 - 1 \times \pi]$
\item [] epochs : $\epsilon = 40$
\item [] stride : s = int
\item [] filter width : w = int
\item [] number of filters : nF = int
\item [] number of layers : nL = int
\item [] qubit count : qC = int
\item [] number of workers : w = int
\item [] Dataset: $(X | Class = \omega)$ 
\item []

\STATE Define circuit compiler with circuit bank cB

\FOR{$\zeta \in \epsilon$}
\STATE Start epoch timer
    \FOR {$nF$}
        \FOR{$x_{k} \in X$}
            \STATE Encode data into unitary matrices $log_n$ encoding
        \STATE Feed filter of size ($w*w$) to CNN
        \STATE Flatten data and run through a dense layer
        \STATE $y(x) = W^{(N)T} h^{(N-1)} + b^{(N)}$
            
            \FOR{$\theta \in \theta_{d}$} 
                \STATE Load $x_{k} \xrightarrow[\mathrm{Data Encoding}]{\mathrm{Quantum}} Q_{Q_{1}} \rightarrow Q_{count} $  
                \STATE Load $\theta_{d} \xrightarrow[\mathrm{Data Encoding}]{\mathrm{Quantum}} Q_{\frac{Q_{count}}{2}+1} + 1  \rightarrow \frac{Q_{count}}{2}+1 $ 
                \STATE Add $\frac{\pi}{2} \rightarrow \theta$ 
                \STATE $\Delta_{fwd} = (E_{Q_{0}}f(\theta_{d}))$
                \STATE Add circuit to cB
                \STATE Subtract $\frac{\pi}{2} \rightarrow \theta$ 
                \STATE $\Delta_{bck} = (E_{Q_{0}}f(\theta_{d}))$
                \STATE Add circuit to cB
            \ENDFOR
            \FOR{$Circuit \in cB$} 
                \STATE Result = Algorithm2(Circuit)
                \STATE Compile list of results from each circuit
            \ENDFOR
        \ENDFOR
    \ENDFOR
    \STATE Stop epoch timer
\ENDFOR
\STATE Record times per epoch
\STATE Calculate accuracy results per epoch
\end{algorithmic}
\label{alg:dist}
\end{algorithm}


\subsection{Distributed Quantum Learning}

The training process of the system is summarized by Algorithm \ref{alg:dist}. 
First, the data is loaded and the user introduces of the training parameters set at run time (Lines 1-2). The learning rate, $\alpha$, indicates how large the updates to the system parameters should be during training. The network weights are initialized randomly. The number of epochs, $\epsilon$, indicates how many times the network will be trained on the data set, $X$. The inputs of the convolutional neural network that factor into the circuit creation such as the number of states, layers, qubits and filters are also initialized.

Next, the circuit compiler where the circuit bank will be stored is initialized and the epoch begins (Lines 3-4). At the start of each epoch an epoch timer is initialized (Line 5). In each epoch, for the number of filters (Line 6), Lines 7-23 must be completed. For each data point in the $X$ (Line 7), the data is encoded into a unitary matrix using $log_n$ encoding (Line 8). From there, a filter shape is determined by the filter width times the width and is fed into the CNN (Line 9). This data is then flattened and run through a classical dense layer (Line 10). Line 11 represents the equation for propagating through the classical dense layer.

Lines 12-22 represent the quantum backpropagation that occurs for each parameter $\theta$. First, the output of the classical layer and all the trainable quantum circuit parameters $\theta_d$ are loaded (Lines 13-14). Then, one forward shifted (Lines 15-17) and one backward shifted (Lines 19-20) circuit are added to the circuit bank. Once the circuit bank is complete, each circuit within the circuit bank is sent to the co-Management Module (Lines 21-22). The results from each circuit execution are retrieved and stored in a list of results (Line 23).

Lastly, the epoch timer is stopped (Line 24) and the time for that epoch is recorded (Line 25). The accuracy results are then calculated and are used to calculate gradients and update the quantum parameters (Line 26).


\subsection{co-Manager in Multi-tenant Quantum System}

The \sol~framework divides a training task into several subtasks, which can be executed independently. This distributed approach effectively leverages the capabilities of multiple quantum machines within the system to improve system performance. However, it also introduces new challenges from a system management perspective. For a given subtask, there may be multiple quantum workers available, necessitating the selection of the optimal worker based on their runtime status.
Given the limited access to quantum hardware and the absence of root privileges, we introduce our quantum-classical co-Manager in a simulated environment. This setup consists of a classical manager and a set of quantum workers represented by simulators. Each quantum worker is equipped with a configurable maximum number of qubits, such as 5 or 10 qubits. The manager is responsible for dynamically managing the quantum workers at runtime. For instance, it must handle join requests from quantum workers wishing to enter the system and assign circuits to the most suitable worker based on their status.
In our system, we delineate four management modules for our co-Manager: (1) co-Manager Initialization; (2) Quantum Worker Registration; (3) Periodic Worker Management; (4) Workload Assignment. Algorithm~\ref{alg:worker} presents our management algorithms.

During the initialization phase, the co-Manager contacts each quantum worker, assigning them IDs (e.g., $w_1, w_2,...,w_n$) in the active worker set $W$. Additionally, it generates a dictionary to record the maximum qubit count, $MR_{w_i}$, for each worker. These values are reported by the workers themselves and stored in the configuration file. Moreover, the co-Manager maintains two other key values for each quantum worker: the available qubits $AR_{w_i}$ and occupied qubits $OR_{w_i}$. These values are updated dynamically based on the runtime workloads on each worker. Finally, the co-Manager maintains other parameters and objects, such as client-submitted circuits, resource demands of the submitted circuits, and active circuit sets on the workers.

During runtime, new workers can dynamically join the system. Worker $w_i$ must register with the co-Manager to accept circuit assignments from the system. In the new worker registration phase, the co-Manager adds $w_i$ to its active worker set, $W$. Since no active circuit executions are taking place on this worker at this point, its occupied qubits $OR_{w_i}$ are set to 0, and its available qubits $AR_{w_i}$ equal its maximum value, $MR_{w_i}$. Furthermore, the co-Manager collects $w_i$'s classical resource usage data (e.g., CPU/GPU), denoted as $CRU_{w_i}(t)$, for further processing (Lines 2-6).

Periodically, the co-Manager must update the runtime status of each quantum worker. This communication is facilitated through heartbeat messages sent from the workers to the co-Manager. Initially, the manager receives the active circuit set, $AC_{w_i}$, from $w_i$ and computes the resource demands for all $c_i \in AC_{w_i}$. The sum represents the occupied qubits, $OR_{w_i}$, at that moment (Lines 7-9). Subsequently, the co-Manager calculates the available qubits (Line 10) and retrieves the current classical resource usage on $w_i$ (Line 11). It is worth noting that the 1 in $t+1$ represents a time unit, which is a configurable value in our system. If the co-Manager fails to receive three consecutive heartbeat messages, it assumes that the system has lost $w_i$ and removes it from the active worker set, $W$ (Lines 12-13).

Since our system comprises multiple quantum workers with varying configurations (e.g., maximum qubits) and runtime statuses (e.g., available qubits and classical resource usage), the co-Manager must identify the best workers for each pending circuit. For a specific circuit, $c_i$, the co-Manager queries the active worker set to find workers with more available qubits than the resource demands of $c_i$. These qualified workers are grouped into a Candidates set (Lines 14-17). When the Candidates set contains multiple workers, the manager sorts them in ascending order based on their most recent classical resource usage (Lines 18-19). In doing so, the co-Manager distributes computational tasks (e.g., circuit execution) according to the workers' runtime statuses, aiming to balance workloads across the system. Eventually, the worker with lowest resource usage is selected to execution $c_i$ (Line 20).

\begin{algorithm}[ht]
\caption{co-Management Modules}
\label{alg:worker}
\begin{algorithmic}[1]
\STATE co-Manager  Initialization:
\item[] Quantum Workers: $W_1, W_2,...,W_n \in W$;
\item[] $MR_{w_i}$: Maximum Resource on worker $i$;
\item[] $AR_{w_i}$: Available Resource on worker $i$;
\item[] $OR_{w_i}$: Occupied Resource on worker $i$;
\item[] $c_i$: Pending circuit $i$;
\item[] $D_{c_i}$: Resource Demand (e.g., qubits) of circuit $i$. 
\item[] $sys_{w_i}$: System call to query current resource usage on $w_i$ 
\item[] $CRU_{w_i}(t)$: Classical Resource Usage on $w_i$ at time $t$;
\item[] $AC_{W_I}$: Active circuits set on $w_i$; 

\item[] 
\STATE New Worker Registration at time $t$: 
\STATE $w_i$ joins $W$;
\STATE $OR_{w_i} = 0$;
\STATE $AR_{w_i} = MR_{w_i}$;
\STATE $CRU_{w_i}(t) = sys_{w_i}$;

\item[] 
\STATE Periodical Heartbeats from $w_i$ at $t+1$:
\FOR{$c_i \in AC_{w_i}$}
    \STATE  $OR_{w_i} = OR_{w_i} + D_{c_i}$
\ENDFOR    
\STATE $AR_{w_i} = MR_{w_i} - OR_{w_i}$;
\STATE $CRU_{w_i}(t+1) = sys_{w_i}$;
\IF {Missing $w_i$'s heartbeats for three periods}
\STATE $w_i$ removed from $W$;
\ENDIF

\item[] 
\STATE Pending circuit $c_i$ assignment at time $t+1$: 
\FOR{$w_i \in W$}
    \IF{$AR_{w_i} > D_{c_i}$}
        \STATE $w_i \rightarrow$ Candidates 
    \ENDIF    
\ENDFOR

\FOR{$w_i \in$ Candidates}
    \STATE Sort ascending based on $CRU_{w_i}(t+1)$; 
\ENDFOR

\STATE Return $w_1$ in Candidates set;

\end{algorithmic}
\end{algorithm}

%% file: result.tex
\section{\sol~ Evaluation}


\subsection{Workload, Implementation and Experiment Settings}

{\noindent\bf Workload}: 
Drawing from previous work on QuClassi~\cite{stein2022quclassi}, we employ quantum-classical convolutional neural networks as our workload. Specifically, it utilizes three variational quantum layer configurations: (1) Single Qubit Unitary, which employs single qubit rotation gates such as $R_y$ and $R_z$ gates; (2) Dual Qubit Unitary, which incorporates two-qubit rotation gates, like $R_{yy}$ and $R_{zz}$; and (3) Entanglement Unitary, which uses two-qubit controlled operations, including $CRY$ and $CRZ$. Although QuClassi serves as our workload, the proposed methodology can be readily adapted to other quantum deep learning applications.

{\noindent\bf Implementation}:
Python 3.9 and the IBM Qiskit Quantum Computing simulator package were used to implement our quantum-classical system. 
The quantum neural network construction, e.g., circuit designs, is adopted from the open-source code of QuClassi~\cite{stein2022quclassi}. The communication between Classical Manager and Quantum Workers is implemented as remote procedure calls through python RPyC library. In our system, the heartbeat messages are exchanged every 5 seconds. This value can be easily configured.

{\noindent\bf Experiment Settings}:
In our workload, we use two qubit settings: 5 qubits and 7 qubits ($qC$ in Algorithm~\ref{alg:dist}), which represent varying circuit widths and computational intensities. Moreover, our circuits feature different numbers of layers: (1) a single layer with Single Qubit Unitary; (2) two layers containing Single and Dual Qubit Unitary; and (3) three layers, encompassing Single, Dual, and Entanglement Unitary. The varying layer counts signify different circuit depths. 
As a result, the workload comprises 1/2/3 layers ($nL$ in Algorithm~\ref{alg:dist}) for both 5-qubit and 7-qubit configurations. 
By running experiments for differing layer and qubit counts, we were able to record an abundance of data and observe a common trend among the results as the number of workers was changed. For the filters, the filter stride was set to 2 pixels ($s$ in Algorithm~\ref{alg:dist}), filter width to 4 pixels ($w$ Algorithm~\ref{alg:dist}) and and the number of filters was set to 4 ($nF$ in Algorithm~\ref{alg:dist}). These settings allowed for images small enough that they could be processed by the lower qubit count computers. Our learning rate was set to 0.001 ($\alpha$ in Algorithm~\ref{alg:dist}).
For training data, we utilize the widely-used MNIST dataset~\cite{mnist}, a handwritten digit classification dataset popular within the quantum deep learning research community.

We conduct experiments on IBM-Q simulation backends, a cloud environment provided by the IBM-Q platform for quantum simulation. However, we lack control over computing resources, such as CPU/GPU cores. Therefore, it is an {\bf uncontrolled environment} to our system. Additionally, we perform experiments on the Google Cloud Platform. By using this commercial cloud, we build our quantum-classical system in a {\bf controlled environment}, specifying the number of CPU cores.
During our experiments, we initially concentrate on classification accuracy to validate the effectiveness. More importantly, we examine completion time and runtime performance to highlight the benefits that applications derive from distributed quantum systems.


\subsection{Accuracy}

As our distributed application is a quantum deep learning-based classifier, we initially compare its accuracy with the non-distributed version to verify its effectiveness. We conducted experiments identical to those described in QuClassi~\cite{stein2022quclassi} using our distributed design and two quantum workers. The results demonstrate the effectiveness of \sol. For instance, in 3/9, 3/8, 3/6, and 1/5 classifications, it achieves accuracies of 97.5\%, 96.2\%, 98.1\%, and 98.6\%, respectively. These accuracies exhibit a difference of less than 2\% when compared to the non-distributed design.

\subsection{Runtime}
In this subsection, we concentrate on runtime evaluation. In these experiments, the client submits its training job to the system, which trains the neural network in a distributed manner. Unlike the previous experiments presented, the runtime experiments do not complete the training; instead, we terminate them upon the completion of one epoch.

\subsubsection{IBM-Q Cloud Backends (Uncontrolled Environment)}

Firstly, we carry out simulations using IBM-Q Cloud Backends. On the classical side, these experiments begin on a local computer—a 2015 Macbook Air—which submits the logical circuits to multiple IBM-Q Cloud Backends. This setup simulates multiple unrestricted quantum workers, i.e., without maximum qubit constraints. The IBM-Q platform simulates the executions and returns the results to our local computer for further processing. This loopback iterates multiple times until the epoch finishes. We conduct experiments with both 5-qubit and 7-qubit configurations.



\begin{figure}[ht]			
	\centering
 \begin{subfigure}[b]{0.24\textwidth}
 	\centering
\includegraphics[width=\textwidth]{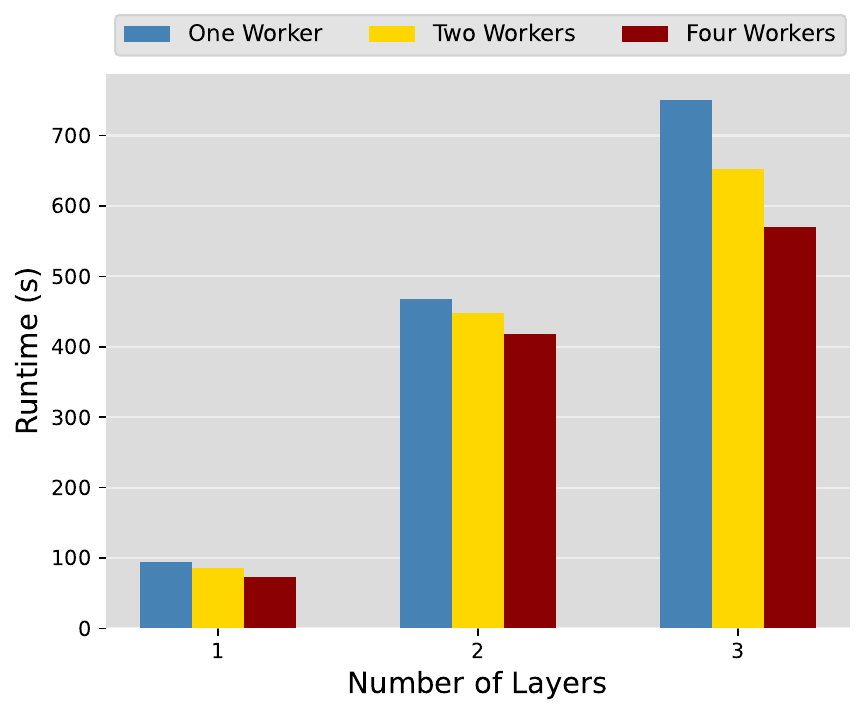}
    \caption{Runtime}
    \label{fig:5Q}
\end{subfigure}
\begin{subfigure}[b]{0.24\textwidth}
	\centering
    \includegraphics[width=\textwidth]{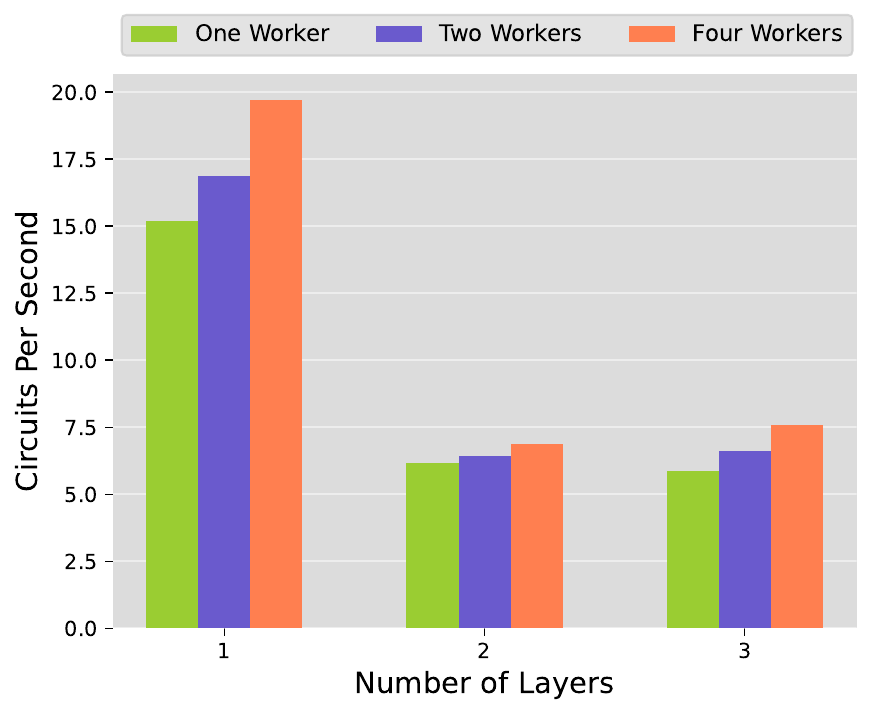}
    \caption{Circuits Per Second}
    \label{fig:5QCirc}
\end{subfigure}
\caption{Experiments on 5-Qubit IBM-Q Backends}
\end{figure}

{\noindent\bf 5-Qubit Setting}:
The data for 5-qubit experiments on circuits with one, two, and three layers can be seen in Figure \ref{fig:5Q}. For all layer counts, the trend is the same. As the number of workers increases, the runtime of the epoch decreases. 

In the one layer experiments, the runtime starts at 94.7 seconds when running on one worker and improves to only 73.1 seconds when using four workers. When using two layers, the recorded runtime for using one simulator is 467.9 seconds and decreases to 418.6 seconds when using four workers. The biggest improvement in terms of the amount of seconds being eliminated is seen in the three layer experiments. When running on one worker the runtime is 749.8 seconds. The time then decreases to 651.7 seconds for two workers and 569.8 seconds for four workers. This is an improvement in runtime of 98.0 and 81.9 seconds, respectively.

Next, we study the number of circuits being run per second based on the number of workers being used. For the various experiments, the number of circuits run are 1440 for one layer, 2880 for two layers, and 4320 for three layers. As seen in Figure \ref{fig:5QCirc}, across all experiments, as the number of workers increases, the number of circuits simulated per second increases. 
The one layer experiment displays the biggest increase as the number of circuits per second jumps from 15.2 when using one worker to 16.9 when using two workers. This number increases to 19.7 when using four workers. That's an overall increase of 4.5 circuits per second when using four workers. The one layer circuits are the least complex circuits used in the experiments so it makes sense that many of them can be simulated per second and the amount per second would increase significantly as the number of workers increases.

For the more complex circuits of two and three layers, the improvement is not as drastic as the one layer experiments, but there is still a steady improvement as the number of workers is increased. During the two layer experiments, the number of circuits per second increases from 6.2 when using one worker to 6.4 when using two workers. This number increases to 6.6 when using four workers. Similar improvement can be seen with the three layer experiments as the number of circuits per second increased from 5.9 with one worker to 6.6 with two workers to 7.6 with four workers. That's an overall increase of 1.7 circuits per second.

\begin{figure}[ht]			
	\centering
 \begin{subfigure}[b]{0.24\textwidth}
 	\centering
\includegraphics[width=1\textwidth]{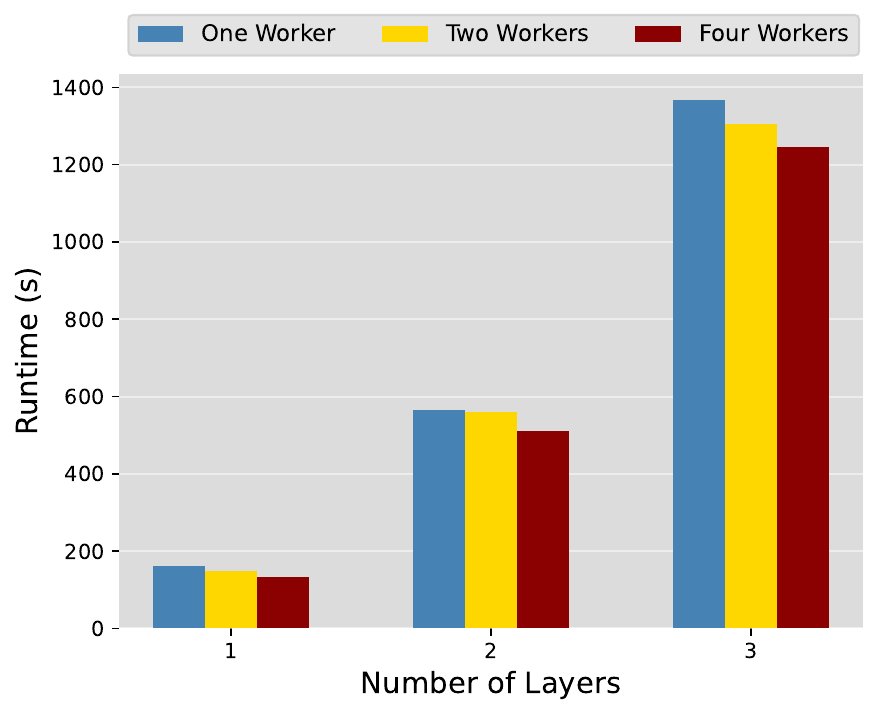}
    \caption{Runtime}
    \label{fig:7Q}
\end{subfigure}
	\centering
\begin{subfigure}[b]{0.24\textwidth}
	\centering
\includegraphics[width=1\textwidth]{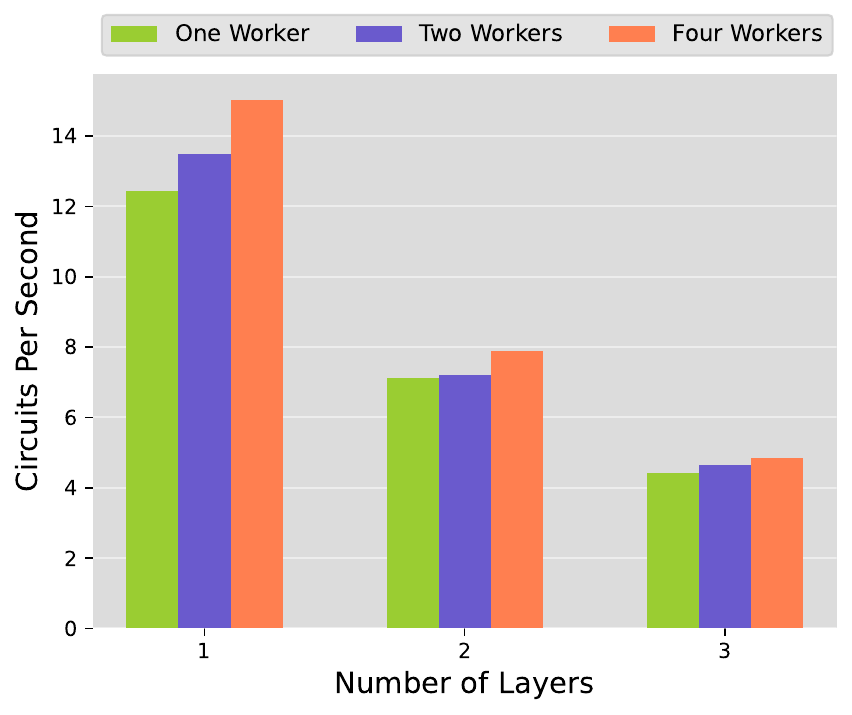}
    \caption{Circuits Per Second}
    \label{fig:7QCirc}
\end{subfigure}
\caption{Experiments on 7-Qubit IBM-Q Backends}
\end{figure}

{\noindent \bf 7-Qubit Setting}:
The data for seven-qubit experiments on circuits with one, two, and three layers can be seen in Figure \ref{fig:7Q}. The results are in line with those of the five-qubit experiments. Each time, more workers were used, the less time it would take for each epoch to run. 
In the one layer experiments, the runtime starts at 163.0 seconds when running on one worker and improves to just 134.3 seconds when using four workers. When using two circuit layers, the recorded runtime for using one worker is 566.5 seconds and decreases to 510.8 seconds when using four workers. The biggest improvement for number of seconds removed once again is seen in the three layer experiments. When running on one worker, the runtime is 1366.1 seconds. The time then decreases to 1303.9 seconds for two workers and 1246.5 seconds for four workers. This is an improvement in runtime of 62.2 and 57.5 seconds, respectively. 

For the seven-qubit experiments, the number of circuits run are 2016 for one layer, 4032 for two layers, and 6048 for three layers. As seen in Figure \ref{fig:7QCirc}, the same trend as the 5-qubit experiments occurs. Across all experiments, as the number of workers increases, the number of circuits simulated per second increases. 
The one layer experiment once again has the largest increase as the number of circuits per second jumps from 12.4 when using one worker to 13.5 when using two workers. This number then increases to 15.0 when using four workers. This is an overall increase of 2.6 circuits per second.
During the two layer experiments, the number of circuits per second slightly increases from 7.1 when using one worker to 7.2 when using two workers. This number increases to 7.9 when using four workers. Improvement can also be seen with three layer experiments as the number of circuits per second increased from 4.4 to 4.6 and to 4.8 with 1-, 2- and 4-workers. 

\subsubsection{Controlled Computing Environment}

In the next step, we deploy our multi-tenant quantum system on the Google Cloud Platform, which provides a controlled computing environment, enabling us to specify the classical hardware configurations. We use e2-medium virtual instances located in the US-Northeast data center. These VMs are configured with 1-2 vCPUs (1 shared core on an Intel Broadwell processor) and 4 GB of memory. In these experiments, we construct 1-manager and 1, 2, and 4-quantum-worker computing environments using the e2-medium instances.
In addition, we conduct two set of experiments, One Client Multiple Circuits and Multiple Clients Multiple Circuits. 

{\noindent\bf One Client Multiple Circuits}: In this setting, one client submits its training job to our system. This job contains multiple circuits that can be executed distributively.  
We configure our quantum workers to the same qubit value, 5 or 7, depending on the experiments. 

\begin{figure}[ht]			
	\centering
 \begin{subfigure}[b]{0.24\textwidth}
 	\centering
\includegraphics[width=1\textwidth]{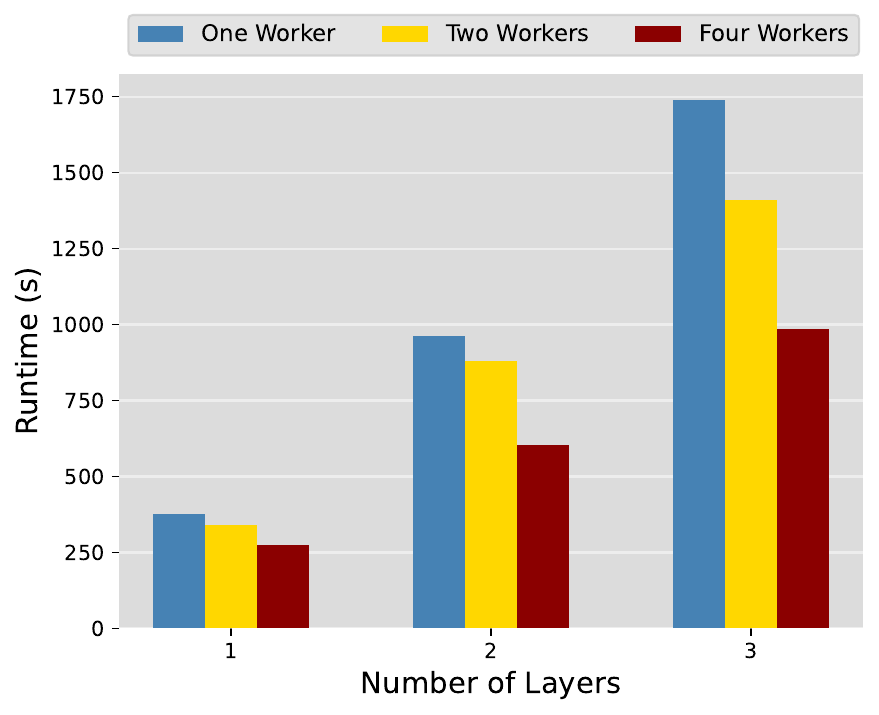}
    \caption{Runtime}
    \label{fig:5q-controlled-runtime}
\end{subfigure}
	\centering
\begin{subfigure}[b]{0.24\textwidth}
	\centering
\includegraphics[width=1\textwidth]{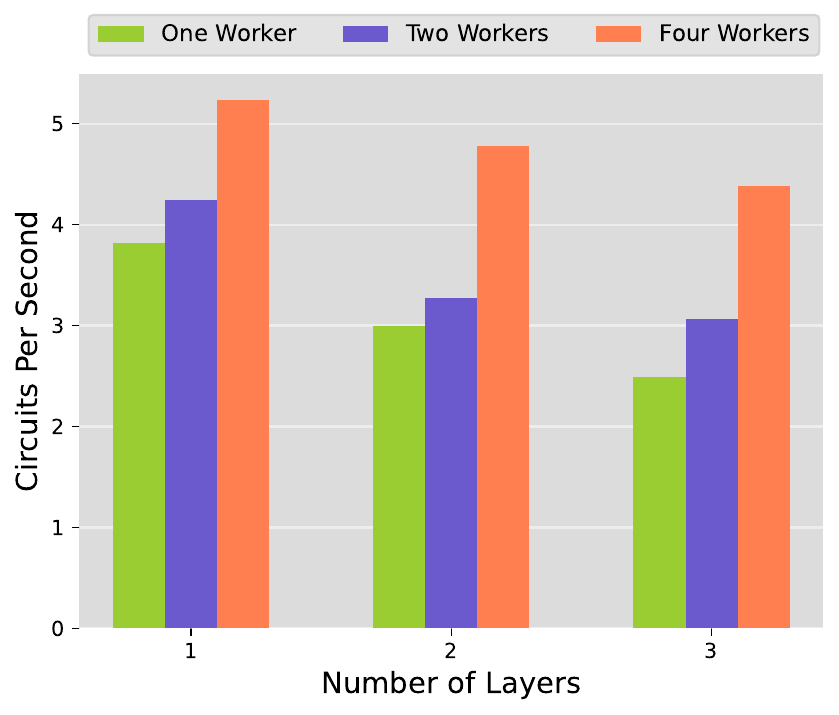}
    \caption{Circuits Per Second}
    \label{fig:5q-controlled-per-second}
\end{subfigure}
\caption{Experiments in a 5-Qubit Controlled Environment}
\end{figure}

Figure~\ref{fig:5q-controlled-runtime} displays the runtime for 5-qubit experiments with 1, 2, and 3 layer settings. As expected, the 4-quantum-worker system demonstrates the shortest runtime. This is primarily due to the presence of 4 quantum workers, allowing the application to distribute its circuits among different workers for execution. The results are then sent back and merged at the client's side. Specifically, the 4-worker system outperforms the 1-worker and 2-worker systems by 27.1\%, 18.9\% in the 1-layer setting, 37.3\%, 31.5\% in the 2-layer setting, and 43.2\%, 30.0\% in the 3-layer setting. It is worth noting that a system with two quantum workers does not reduce the runtime of the one-worker system by half. This is because, as the number of workers increases, the overhead, such as communication costs and quantum state analysis, also grows.

Upon closer examination of the experiments, it becomes evident that the distributed system with multiple workers processes more circuits per second. Figure~\ref{fig:5q-controlled-per-second} presents the results. In the 1-layer setting, the processing speed is 3.8, 4.2, and 5.2 circuits per second for the 1-worker, 2-worker, and 4-worker systems, respectively. This represents a 36.8\% increase from the 1-worker to 2-worker system and a 23.8\% increase from the 2-worker to 4-worker system. Examining the more computationally intensive 3-layer setting, the values are 2.4, 3.1, and 4.4 circuits per second, reflecting gains of 83.3\% and 41.9\% as the number of workers increases.

\begin{figure}[ht]			
	\centering
 \begin{subfigure}[b]{0.24\textwidth}
 	\centering
\includegraphics[width=1\textwidth]{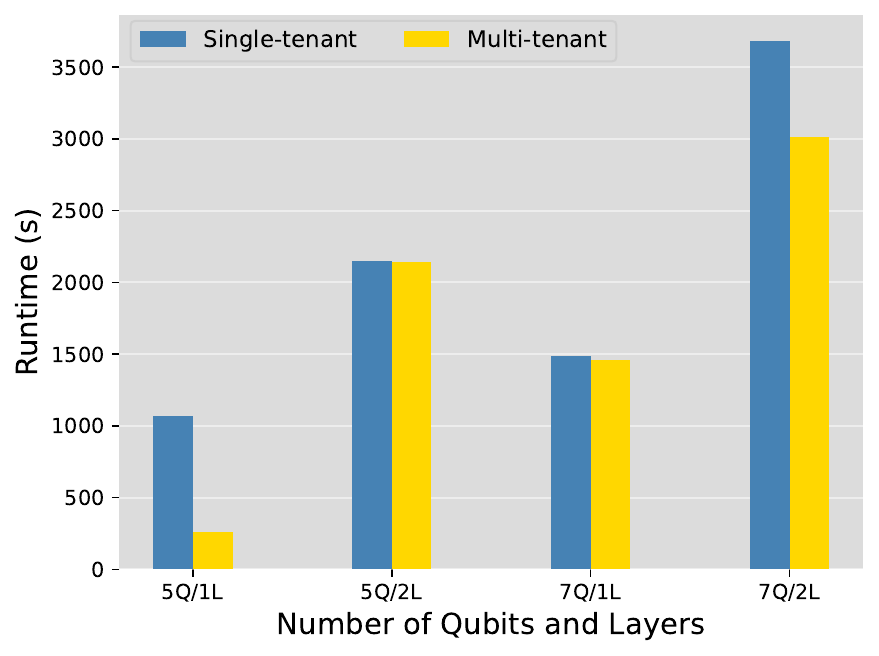}
    \caption{Runtime}
    \label{fig:concurrent-controlled-runtime}
\end{subfigure}
	\centering
\begin{subfigure}[b]{0.24\textwidth}
	\centering
\includegraphics[width=1\textwidth]{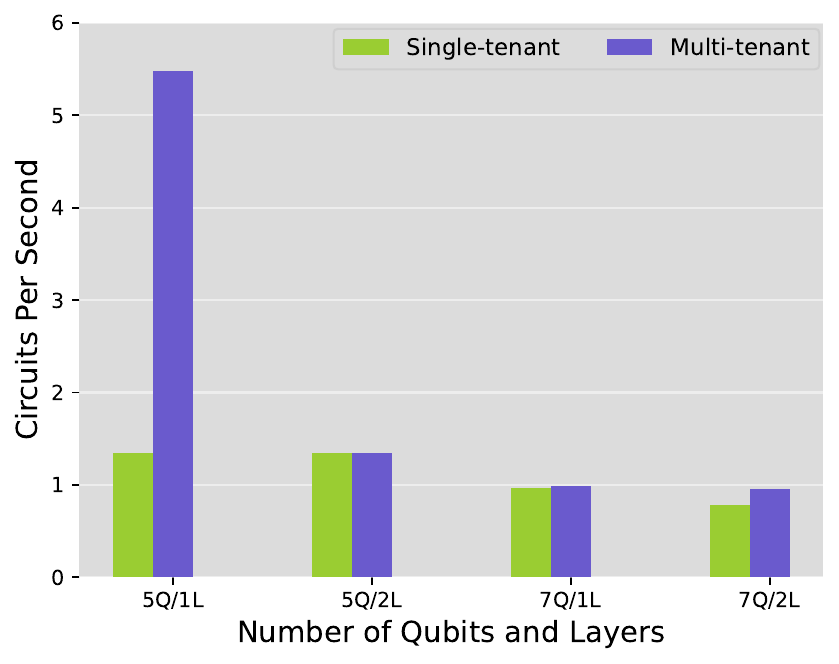}
    \caption{Circuits Per Second}
    \label{fig:concurrent-controlled-circuit-per-second}
\end{subfigure}
\caption{Experiments in a Multi-tenant System}
\end{figure}

{\noindent\bf Multi Clients Multiple Circuits}: 
In this setting, we have four concurrent clients, each submitting a training application, such as 5-qubits-1-layer, which contains multiple distributed circuits. We launch four VMs to represent four quantum workers and configure them with 5, 10, 15, and 20 qubits. As our circuits are either 5 or 7 qubits in width, a 20-qubit machine can accommodate four 5-qubit circuits and two 7-qubit circuits concurrently. Depending on the co-Manager's workload distribution, it may also execute two 7-qubit and one 5-qubit circuits simultaneously.
Figure~\ref{fig:concurrent-controlled-runtime} presents the runtime results. When compared to the results from Figure~\ref{fig:5q-controlled-runtime}, we can see that the completion time has increased significantly. This is due to the fact that the previous experiment contains only one job with multiple distributed circuits. In contrast, this experiment includes multiple concurrent jobs, leading to a substantial increase in the computing workload.

Compared to a single-tenant system, the multi-tenant system achieves significant runtime reductions of 68.7\% and 8.2\% for 5Q/1L and 7Q/2L, respectively. This is because, in general, circuits can share the same worker in a multi-tenant distributed system, leading to a more efficient use of system-wide resources. Specifically, at the very beginning, the system has workloads, allowing circuits from 5Q/1L to be distributed to 20-qubit workers for faster processing. Furthermore, at the end of the experiment, as the other three jobs have been completed, the system becomes less congested, making more resources available for 7Q/2L. Comparing 5Q/1L and 7Q/2L both in a multi-tenant system, the improvement reduces due to the fact that quantum worker-1, which only has 5 qubits, is useless to a 7-qubit circuit. 
However, the 5Q/2L and 7Q/2L have similar performance. This is because, during their runtime, the system is highly congested with all four training jobs, leaving less room for optimization within our system.

Figure~\ref{fig:concurrent-controlled-circuit-per-second} displays the circuits per second data from the concurrent clients. The same trend is observed, with the 5-qubit-1-layer achieving the largest gain of 3.9 times, increasing from 1.4 to 5.5. When compared to Figure~\ref{fig:5q-controlled-per-second}, the degree of improvement increases significantly due to the multi-tenancy enabled in the system, allowing concurrent clients to share the same quantum worker.

%% file: discussion.tex
\section{Discussion and Conclusion}

In this project, we investigate \sol, a distributed quantum learning approach within a quantum-classical system featuring a manager-worker architecture. 
Our distributed design aims to fully utilize all available quantum workers across the system. 
We propose co-Management modules that efficiently manage the quantum workers within the system and dynamically distribute circuits to the most suitable workers.
We conduct experiments in both uncontrolled and controlled computing environments, namely IBM-Q Cloud Backend and Google Cloud Platform. The results demonstrate the effectiveness of the \sol~system, achieving similar accuracies as its non-distributed counterpart. Furthermore, it exhibits substantial reductions in runtime, by up to 68.7\%, and significant improvements in processed circuits per second, with gains of up to 3.9 times, in a 4-worker multi-tenant setting.

Despite the advantages obtained, the system has some limitations. Firstly, the communication between a classical manager and quantum workers relies solely on classical channels (e.g., remote procedure calls), failing to exploit the potential of quantum networking, 
Secondly, our system does not take noise into account when scheduling the workload. However, quantum noise has a significant impact on state fidelities. Lastly, due to the lack of privileges on quantum hardware, our evaluations are limited to simulations.


\section*{Acknowledgement}
This work is supported in part by the National Science Foundation (NSF) under grant agreements 2329020, and 2301884. This material is also based upon work supported by the U.S. Department of Energy, Office of Science, National Quantum Information Science Research Centers, Co-design Center for Quantum Advantage (C2QA) under contract number DE-SC0012704, (Basic Energy Sciences, PNNL FWP 76274).